\documentclass[journal]{IEEEtran}

\usepackage{xcolor,soul,framed} 

\colorlet{shadecolor}{yellow}
\usepackage[pdftex]{graphicx}
\graphicspath{{../pdf/}{../jpeg/}}
\DeclareGraphicsExtensions{.pdf,.jpeg,.png}

\usepackage[cmex10]{amsmath}
\usepackage{array}
\usepackage{mdwmath}
\usepackage{mdwtab}
\usepackage{eqparbox}
\usepackage{url}
\usepackage{xcolor}
\usepackage{soul}
\usepackage{subfigure}
\usepackage{graphicx}
\usepackage{float}

\begin{document}
\bstctlcite{IEEEexample:BSTcontrol}
    \title{Analytical method for the diffraction of an electromagnetic wave by subwavelength graphene ribbons}
  \author{Mahdi~Rahmanzadeh,~Amin~Khavasi ,and~Behzad~Rejaei

    \thanks{The authors are with the Department of Electrical Engineering, Sharif University of Technology, Tehran 11155-4363, Iran (email:rahmanzadeh.mahdi@ee.sharif.edu; khavasi@sharif.edu; rejaei@sharif.edu}}

\maketitle

\begin{abstract}
Theoretical study of arrays of graphene ribbons is currently of high interest due to its potential application in beam splitters, absorbers, and polarizers. In this paper, an analytical method is presented for diffraction analysis of graphene ribbon arrays. Previous analytical studies were carried out in the regime where the lateral separation between the ribbons is much smaller than the wavelength of the incident wave. As such, they could not be used to calculate the reflection coefficients of higher diffracted orders. By contrast, the method proposed here can predict electromagnetic response of graphene ribbon arrays even when the array constant is larger than the wavelength. To reach our results, we first derive an analytical expression for the surface density of electric current induced on the ribbons by an incident, transverse-magnetic (TM), plane wave. Next, closed-form and analytical expressions are obtained for the reflection coefficients of the zeroth and higher diffracted orders. The results are in excellent agreement with those obtained from full-wave simulations. The method presented facilitates the design of tunable gratings with many potential applications in THz and optics.
\end{abstract}

\section{Introduction}

\IEEEPARstart{I}{N} the past decade, the diffraction analysis of periodic photonic structures and gratings has attracted a great deal of interest due to  numerous applications in spectroscopy, imaging, anomalous reflectors, and external cavity lasers\cite{kneubuhl1969diffraction,harvey1991external,chrisp1999convex,blanchard1999simultaneous,jull1980unusual,ra2017metagratings,rabinovich2018analytical,rahmanzadeh2020perfect}. In particular, tunable diffraction gratings have been investigated for instantaneous control over the diffraction pattern. The tunability of the structure can be attained by mechanically changing or external biasing voltages \cite{tian2017electronically,ra2018reconfigurable,casolaro2019dynamic,li2006tunable}. 

At THz frequencies, graphene-based structures may be used for realization of tunable diffraction gratings \cite{ra2018reconfigurable,wang2017tunable,yan2019ultrasensitive,patel2018graphene,behroozinia2020real}. Graphene, a two-dimensional material made of carbon atoms and arranged in a honeycomb lattice \cite{geim2010rise}, is characterized by high electrical conductivity and controllable plasmonic properties. The surface conductivity of graphene can be changed by varying the electrostatic bias of the material. These properties make graphene a promising material for various applications ranging from perfect absorption to tunable wavefront shaping \cite{mak2008measurement,grigorenko2012graphene,novoselov2004electric,bao2012graphene,rahmanzadeh2018multilayer,momeni2018information,chen2013nanostructured,chamanara2013non,hosseininejad2019digital,rajabalipanah2020real,andryieuski2013graphene}.
In \cite{ra2018reconfigurable}, metallic/graphene hybrid strips were suggested for realizing reconfigurable gratings whose characteristics can be tuned by electrostatically biasing the graphene strips. Moreover, in \cite{behroozinia2020real} , multifunctional tunable gratings were proposed for operating at THz frequencies. The designed grating was composed of three graphene ribbons and could realize multiple functionalities such as an anomalous reflector, retroreflector, and the beam splitter.

 Design and analysis of graphene-based diffraction gratings, like most other graphene-based devices, rely heavily on time-consuming full-wave simulations and  sophisticated design procedures. Numerical methods usually suffer from poor convergence and are extremely time-consuming\cite{khavasi2013fast}. Analytical or semi-analytical treatment  of these structures is, therefore, critical to the understanding of physical principles behind their operation and designing novel devices. In\cite{rahmanzadeh2019analytical,rahmanzadeh2018adopting,barzegar2015analytical,khavasi2014analytical}, accurate and fast analytical solutions were presented to analyze periodic arrays of graphene ribbons and disks. These solutions, obtained from solving  integral equations governing the surface current density on patterned graphene, are precise, affordable, and reliable; hence, several novel devices have been proposed using these methods\cite{abdollahramezani2015beam,arik2017polarization,zhao2017terahertz,khavasi2015design}. However, these solutions were obtained under a quasi-static approximation, valid when the period of the structure is much smaller than the operating wavelength. 
As such, these analytical methods cannot be used to design graphene-based gratings with propagating non-zero diffraction orders.

In this paper, we generalize the work in \cite{khavasi2014analytical} to obtain an analytical method for investigation of arrays of narrow graphene ribbons with an arbitrary array constant. Integral equations governing the surface currents induced on the ribbons shall be derived and  solved by expanding the current density in terms of the eigenfunctions of the problem of a single graphene ribbon. Next, the Rayleigh expansion, in combination with the appropriate boundary conditions, will be used to determine the amplitude of the diffracted orders. The method will be validated against full-wave simulations through some numerical examples. Finally, the limitations of our proposed method are discussed.

\section{Surface currents on a periodic array of graphene ribbons}

\begin{figure}
\centering \includegraphics[width=\linewidth]{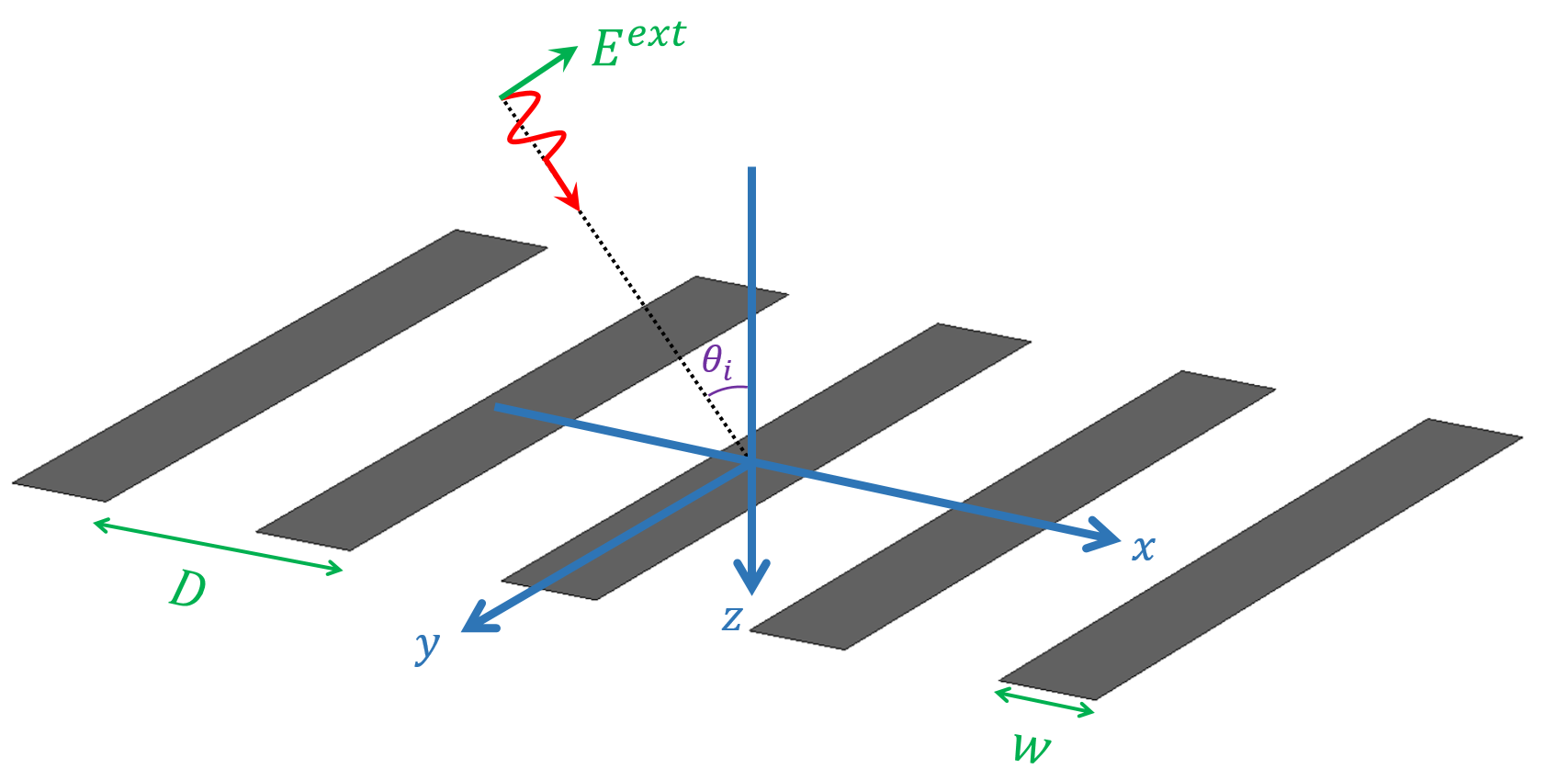}
\caption{Periodic array of graphene ribbons. Graphene ribbons are illuminated by an oblique TM polarized wave with the angle of $\theta_i$}
\label{GR}
\end{figure}

Consider a periodic array of graphene ribbons in free space as shown in Fig.\ref{GR}. The width of the ribbons and the array constant  (along the $x$-axis) are $w$ and $D$, respectively. The structure is infinite along the $y$-axis ($\partial/\partial y\sim 0$). A plane wave with transverse-magnetic (TM) polarization is incident upon the array with an angle of incidence of $\theta_i$. Although the array constant $D$ is arbitrary, it is assumed that $w\ll \lambda$, with $\lambda_0$ the free-space wavelength of the incident wave. 

Graphene is modeled as a surface conductivity $\sigma_s$, computed within the random-phase approximation
\begin{equation}
  \label{Graphene Sigma}
   \begin{split}
\sigma _s = \frac{2e^2 k_B T}{\pi \hbar ^2}\frac{j}{j \tau ^{-1} - \omega }\ln [2\,\cosh(E_F/2 k_B T)] \\ + \frac{e^2}{4\hbar }[H(\omega /2) - \frac{4j\omega }{\pi }\int\limits_0^\infty{\frac{H(\varepsilon ) - H(\omega /2)}{{\omega ^2 - 4 \varepsilon ^2}}d\varepsilon}
\end{split}
  \end{equation}
where $e$ is the electron charge, $E_F$ is the Fermi energy, $\hbar$ is the reduced Plank constant, $k_B$  is the Boltzmann constant, $\omega$ is the frequency, $T$ is the temperature and $\tau$ is the relaxation time. The function $H(\varepsilon)$ is given by
\begin{equation}
  \label{H(eps)}
H(\varepsilon)=\frac{\sinh(\hbar \varepsilon/k_B T)}{\cosh(E_F/k_B T)+\cosh(\hbar \varepsilon/ k_B T)}  \end{equation}
In the following, a time dependence of the form $\exp(j{\omega}t)$ is implicitly assumed. 

Due to uniformity of the structure along the $y$-direction, the surface currents induced on the graphene ribbons by the TM-polarized wave will have a $x$-component only. Because of Floquet theorem, current density on each ribbon (designated by an index $l$) can be related to the current density $J_x(x)$ on a reference ribbon ($l=0$) by 
\begin{equation}\label{floq}
J_{x,l}(x)=e^{-jk_{x}lD} J_{x}(x+lD)
\end{equation}
where 
\begin{equation}
    k_{x}=k_{0}\sin(\theta_{i})
\end{equation}
The electric field on  each ribbon is the sum of the external (incident) electric field and the electric field generated in free space by the induced surface currents on all ribbons. By relating the total electric field to surface current density, and using \eqref{floq}, an integral equation can be obtained for the surface current on the reference ribbon \cite{rahmanzadeh2019analytical}
\begin{equation}
  \label{Main_IE}
    \begin{split}
\frac{{J_x}(x)}{\sigma_s} = E_x^{ext}(x) + \frac{1}{{j\omega {\varepsilon _0}}}\frac{d}{{dx}}\int\limits_{ - w/2}^{w/2} {{G_x}(x - x')} \,\frac{{d\,{J_x}(x')}}{{dx'}}dx' \\ -  j\omega {\mu _0}\int\limits_{ - w/2}^{w/2} {{G_x}(x - x')\,} {J_x}(x')dx'
   \end{split}
  \end{equation}
where $E^{ext}_x (x)$ is the $x$-component of the external electric field, and 
\begin{equation}
  \label{Periodic_GF}
{G_x}(x - x') = \frac{1}{{4j}}\sum\limits_{l =  - \infty }^\infty  {{e^{j{k_x}lD}}H_0^{(2)}({k_0}\left| {x - x' - lD} \right|)} 
 \end{equation}
is the free-space periodic Green's function. In \eqref{Periodic_GF}, $k_0$ is the free space wave number and $H^{(2)}_{0}$ denotes the zeroth order Hankel function of the second kind. The last two terms on the right hand side of \eqref{Main_IE} represent the electric field produced by surface current on the ribbons. 

Instead of directly solving \eqref{Main_IE}, we shall attempt to find an approximate, but accurate solution by first considering the problem of a single ribbon that is subject to the same incident wave. The surface current density $J^{0}_{x}(x)$ will then satisfy the same integral equation \eqref{Main_IE}, but with the periodic Green's function $G_{x}(x-x^{\prime})$ replaced by the ordinary 2D free space Green's function
\begin{equation}
 \label{GF0}
{G_0}(x - x') = \frac{1}{{4j}}  
H_0^{(2)}({k_0}\left| {x - x'} \right|)
 \end{equation}
This is equivalent to retaining the $l=0$ term in the summation in \eqref{Periodic_GF}. It was shown in   \cite{khavasi2014analytical} that for a single, narrow ribbon where $w\ll \lambda_0$ (or $k_0 w \ll 1$), the resulting integral equation may then be approximated by the quasi-static integral equation 
\begin{equation}
  \label{qs}
\frac{{J^{0}_x}(x)}{\sigma_s} = E_x^{ext}(x) - \frac{1}{{2\pi j\omega {\varepsilon _0}}}\int\limits_{ - w/2}^{w/2} \frac{1}{x-x^{\prime}} \,\frac{{d\,{J^{0}_{x}}(x')}}{{dx'}}dx' 
  \end{equation}
Solution is obtained by considering the eigenvalue problem 
\begin{equation}
\frac{1}{\pi}\int_{-w/2}^{w/2}\frac{1}{x-x'}\frac{d\psi_{n}(x')}{dx'}dx'=q^{0}_{n}\psi_{n}(x)\label{eigv0}
\end{equation}
and expanding $J^{0}_{x}(x)$ using the set of real eigenfunctions $\psi_{n}(x)$ that satisfy the orthonormality condition
\begin{equation}\label{ortho}
\int_{-w/2}^{w/2} \psi_{n}(x)\psi_{m}(x) dx = \delta_{nm}
\end{equation}
with $\delta_{mn}$ the Kronecker delta. A method for calculating $\psi_{n}(x)$  and $q_{n}^{0}$ using Fourier expansion  eigenfunctions is presented in \cite{khavasi2014analytical}. The first three eigenfunctions are listed in Table{\ref{eigenfunctions}}. The higher-order eigenfunctions ($n > 3$) can be approximately determined by $\sqrt{2/w} \cos(n\pi x/w)$ and $\sqrt{2/w} \sin(n\pi x/w)$ for odd and even orders, respectively. Solution of \eqref{qs} may be written as
\begin{subequations}
\label{Expan_Jx}
\begin{equation}
  \label{J_x}
{J^{0}_x}(x) = \sum\limits_{n = 1} {{A^{0}_n}{\psi _n}(x)}
  \end{equation}
  \begin{equation}
  \label{A_n}
A^{0}_n=\frac{\sigma_s}{1+ q^{0}_n \sigma_s/(2j\omega\varepsilon_{0})} \int\limits_{ - w/2}^{w/2} {{\psi _n}(x)\,E_x^{ext}(x)dx}
  \end{equation}
\end{subequations}

\begin{table} [ht!]
\caption{The first three eigenfunctions for the problem of a single graphene ribbon.}
\begin{center}
\begin{tabular}{c} 

\hline
 Eigenfunction \\ \hline
${\psi _1} = {w^{ - 0.5}}[1.2\sin \left( {\arccos {\kern 1pt} {{2x} \mathord{\left/
 {\vphantom {{2x} w}} \right.
 \kern-\nulldelimiterspace} w}} \right) - 1.06\sin \left( {3\arccos {{2x} \mathord{\left/
 {\vphantom {{2x} w}} \right.
 \kern-\nulldelimiterspace} w}} \right)] 
$ \\ \hline
 ${\psi _2} = {w^{ - 0.5}}[1.254\sin \left( {2\arccos {\kern 1pt} {{2x} \mathord{\left/
 {\vphantom {{2x} w}} \right.
 \kern-\nulldelimiterspace} w}} \right) - 0.302\sin \left( {4\arccos {{2x} \mathord{\left/
 {\vphantom {{2x} w}} \right.
 \kern-\nulldelimiterspace} w}} \right)]$  \\ \hline
\begin{array}{l}{\psi _3} = {w^{ - 0.5}}\,[0.308\sin \left( {\arccos {\kern 1pt} {{2x} \mathord{\left/
 {\vphantom {{2x} w}} \right.
 \kern-\nulldelimiterspace} w}} \right) + 1.19\sin \left( {3\arccos {{2x} \mathord{\left/
 {\vphantom {{2x} w}} \right.
 \kern-\nulldelimiterspace} w}} \right)\\\,\,\,\,\,\,\,\,\,\, - 0.484\sin \left( {5\arccos {{2x} \mathord{\left/ 
 {\vphantom {{2x} w}} \right.
 \kern-\nulldelimiterspace} w}} \right)]\end{array}

 \\ \hline

\end{tabular}
\end{center}

\label{eigenfunctions}
\end{table}

For an array of ribbons with subwavelength array constant, it was assumed in \cite{khavasi2014analytical} that (i) the  periodic Green's function of the array may be replaced by a quasi-static Green's function and (ii) that the interaction between the ribbons in the array may be taken into account as a perturbation. Here, the first assumption is clearly not valid, as the separation between ribbons is arbitrary. However, we once again apply perturbation theory to find an approximate solution of \eqref{Main_IE}. Let us rewrite this equation as
\begin{align}
  \label{Main_IEo}
\frac{{J_x}(x)}{\sigma_s} &= E_x^{ext}(x) -\frac{1}{2j\omega\varepsilon_{0}}\int\limits_{ - w/2}^{w/2} \hat{\Gamma}(x-x') J_x(x')dx'\\
\hat{\Gamma}(x-x')=& 
-2\frac{d}{dx}G_x(x - x') \,\frac{d}{dx'} - 2 k_{0}^{2}G_x(x - x')\label{gamma}
  \end{align}
In analogy with the problem of a single ribbon, one may next consider the eigenvalue problem 
\begin{equation}
\label{eig}
\int_{-w/2}^{w/2}\hat{\Gamma}(x-x')\phi_{n}(x')dx'=q_{n}\phi_{n}(x)
\end{equation}
and expand $J_{x}$ in the eigenfunctions $\phi_{n}$ of the integral operator $\hat{\Gamma}$. This will result in an expansion identical to \eqref{Expan_Jx} except for $q_{0}^{0},\psi_n$ replaced by $q_n,\phi_n$ in \eqref{A_n}. Instead of trying to solve \eqref{eig}, however, we note that if the interaction between the ribbons is small, one may write the operator $\hat{\Gamma}(x-x^{\prime})$ as 
\begin{equation}\label{delt}
    \hat{\Gamma}(x-x')=\frac{1}{\pi } \frac{1}{x-x^{\prime}} \,\frac{{d}}{dx'} +\Delta \hat{\Gamma}(x-x')
\end{equation}
an consider $\Delta \hat{\Gamma}(x-x')$ as a perturbation. The motivation behind this approach is that, as in \cite{khavasi2014analytical}, the electromagnetic interaction between the ribbons is small in comparison with self-interaction [$l=0$ term in \eqref{Periodic_GF}], and that, for a narrow ribbon, self-interaction is well described by retaining the quasi-static contribution only. 

Considering $\Delta \hat{\Gamma}(x-x')$ as a perturbation, we use first order perturbation theory to approximate $q_{n}$ using the "unperturbed" eigenvalues $q_{n}^{0}$ and eigenfunctions $\psi_{n}$:
\begin{equation}
    q_{n}=q_{n}^{0}+\int_{-w/2}^{w/2}\int_{-w/2}^{w/2}
    \psi_{n}(x)\Delta\hat{\Gamma}(x-x^{\prime})\psi_{n}(x^{\prime}) dx dx^{\prime}
\end{equation}
which, using  \eqref{eigv0}, \eqref{ortho}, and \eqref{delt}, can be written as
\begin{equation}
    q_{n}=\int_{-w/2}^{w/2}\int_{-w/2}^{w/2}
    \psi_{n}(x)\hat{\Gamma}(x-x^{\prime})\psi_{n}(x^{\prime}) dx dx^{\prime}\label{doub}
\end{equation}
Finally, we assume that eigenfunctions are not significantly affected by the perturbation $\Delta\hat{\Gamma}$, and approximate $\phi_n(x)$ by $\psi_n(x)$. This yields the approximate solution for the current density 
\begin{equation}
  \label{Jx2}
{J_x}(x) = \sum\limits_{n = 1} {{A _n}{\psi_n}(x)}
  \end{equation}
where $A_n$ is given by \eqref{A_n}, with $q_{n}^{0}$ replaced by $q_n$: 
\begin{equation}
  \label{A_n2}
A_n=\frac{\sigma_s}{1+ q_n \sigma_s/(2j\omega\varepsilon_{0})} \int\limits_{ - w/2}^{w/2} {{\psi _n}(x)\,E_x^{ext}(x)dx}
  \end{equation}

In order to evaluate $q_{n}$, we  use the following relationship for $G_x(x-x')$ \cite{ansari2020local}
\begin{equation}
  \label{PGF_expan}
G_x(x-x')= \frac{1}{{2jD}}\sum\limits_{p =  - \infty }^\infty  {\frac{{{e^{ - j\left( {{k_x} + \frac{{2\pi p}}{D}} \right)\left( {x - x'} \right)}}}}{{\sqrt {k_0^2 - {{\left( {{k_x} + \frac{{2\pi p}}{D}} \right)}^2}} }}}
  \end{equation}
Substitution in \eqref{gamma},\eqref{doub}, then yields
\begin{equation}
  \label{q-n}
  {q_n} = \frac{1}{{D}}\sum\limits_{p =  - \infty }^\infty  
  \sqrt{k_{xp}^{2}-k_0^{2}} \left|f_{pn}\right|^{2}
  \end{equation}
where
\begin{subequations}
\label{FT}
  \begin{equation}
  \label{F}
{f_{pn}} =\int\limits_{-w/2}^{w/2} 
e^{jk_{x,p}x}\psi_n(x)\,dx 
\end{equation}
\begin{equation}
  \label{k_xp}
{k_{x,p}} = {k_x} + \frac{{2p\pi }}{D}
\end{equation}
 \end{subequations}

\section{Reflection and transmission coefficients of diffraction orders}

\begin{figure}
\centering \includegraphics[width=\linewidth]{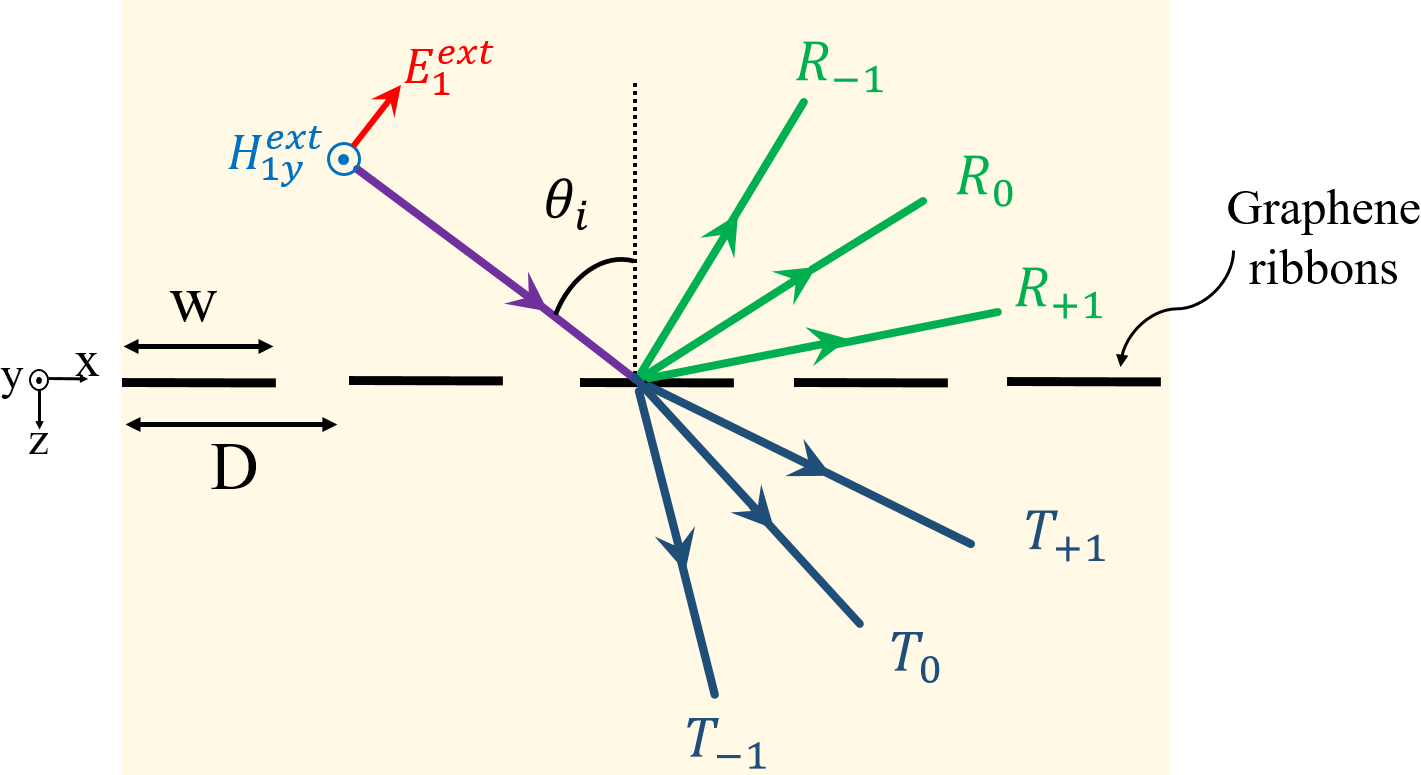}
\caption{Diffraction of an oblique TM-plane wave by an array of graphene ribbons.}
\label{GR_Diffraction}
\end{figure}

The current distribution on graphene ribbons was derived in the previous section. To calculate the reflection coefficient of diffraction orders, we must expand the total field above and below the grating by using Floquet-Bloch theory which states that when a plane wave is scattered by a periodic structure, a discrete set of  waves is diffracted as shown in Fig.\ref{GR_Diffraction}. In case of an incident TM plane wave of unit amplitude, with the wave vector $\mathbf{k}=k_{x}\hat{\mathbf{x}}+k_{z}\hat{\mathbf{z}}, k_{z}=\sqrt{k_{0}^{2}-k_{x}^{2}}$,  The electromagnetic field in the region $z<0$ is given by 
\begin{subequations}
\label{Field_exp_1}
\begin{equation}
  \label{H1y}
{H_{1y}} = {e^{- j{k_{x}x} - j{k_{z}}z}} + \sum\limits_{m =  - \infty }^\infty  {{R_m}{e^{- j{k_{x,m}}x+j{k_{z,m}}z}}} 
 \end{equation}
 \begin{equation}
  \label{E1x}
{E_{1x}} = {\xi _{0}}{e^{- j{k_{x}}x - j{k_{z}}z}} - \sum\limits_{m =  - \infty }^\infty  {{\xi _{m}}{R_m}{e^{- j{k_{x,m}}x+j{k_{z,m}}z}}}
  \end{equation}
\end{subequations}
where $k_{x,m}$ is defined by \eqref{k_xp} and 
\begin{subequations}
\label{kx-kz-zita}
  \begin{equation}
  \label{kz-m}
{k_{z,m}} =  \sqrt{k_{0}^{2}-k_{x,m}^{2}}
\end{equation}
\begin{equation}
  \label{zita-n}
{\xi _{m}} = {k_{z,m}}/(\omega {\varepsilon _0})
  \end{equation}
  \end{subequations}
For  $z>0$,

\begin{subequations}
\label{Field_exp_2}
\begin{equation}
  \label{H2y}
{H_{2y}} = \sum\limits_{m =  - \infty }^\infty {T_m}{e^{ - j{k_{x,m}}x- j{k_{z,m}}z}}
  \end{equation}
  
 \begin{equation}
  \label{E2x}
{E_{2x}} = \sum\limits_{m =  - \infty }^\infty  {{\xi _{m}}{T_m}{e^{ - j{k_{x,m}}x- j{k_{z,m}}z}}}
  \end{equation}
\end{subequations}
In these equations, $R_n$ and $T_n$ denote the reflection, respectively,
transmission coefficient  of $n$-th diffraction order. The $z$ component $k_{z,n}$ of the wave vector of the $n$-th order diffracted wave is either negative real (propagating wave) or positive imaginary (evanescent wave). 

We next apply the electromagnetic boundary conditions at $z=0$,
\begin{subequations}
\label{BC}
\begin{equation}
  \label{E_BC}
E_{1x} = E_{2x}
  \end{equation}
  \begin{equation}
  \label{H_BC}
{H_{1y}} - {H_{2y}} = {J_x}(x)
  \end{equation}
\end{subequations}
Substituting \eqref{Field_exp_1} and \eqref{Field_exp_2} in \eqref{BC}, one obtains, after some straightforward mathematical manipulations,
\begin{subequations}
\label{T0-T_n}
 \begin{equation}
  \label{T_n}
{T_m} =  - {R_m} ,\,\,;\,m \ne 0
  \end{equation}
  \begin{equation}
  \label{T0}
{T_0} = 1 - {R_0}
  \end{equation}
 \end{subequations}
  \begin{equation}
  \label{rj}
\sum\limits_{m \ne 0} {2 {R_m}{e^{ - j{k_{x,m}}x}}} + 2 {R_0}{e^{ - j{k_{x}}x}} = {J_x}(x)
\end{equation}
Multiplying both sides of \eqref{rj} by $e^{jk_{x,m} x}$ and taking the integral over one period, the reflection coefficients are readily obtained as
\begin{equation}
  \label{R_n}
{R_m} = \frac{1}{{2D}}\sum\limits_{n = 1} {{A_n}f_{mn}} 
  \end{equation}
where $A_n$ and $f_{mn}$ were defined by \eqref{A_n2} and \eqref{F}, respectively.

Finally, we define the diffraction efficiencies, i.e. the ratio of the diffracted power to the incident power, as follows

\begin{equation}
  \label{DE}
DE_{m}=R_m R_m^* Re(\frac{k_{z,m}}{k_{z}})  
  \end{equation}

\section{Numerical results and discussion}

\begin{figure} 

\centering\includegraphics[width=\linewidth]{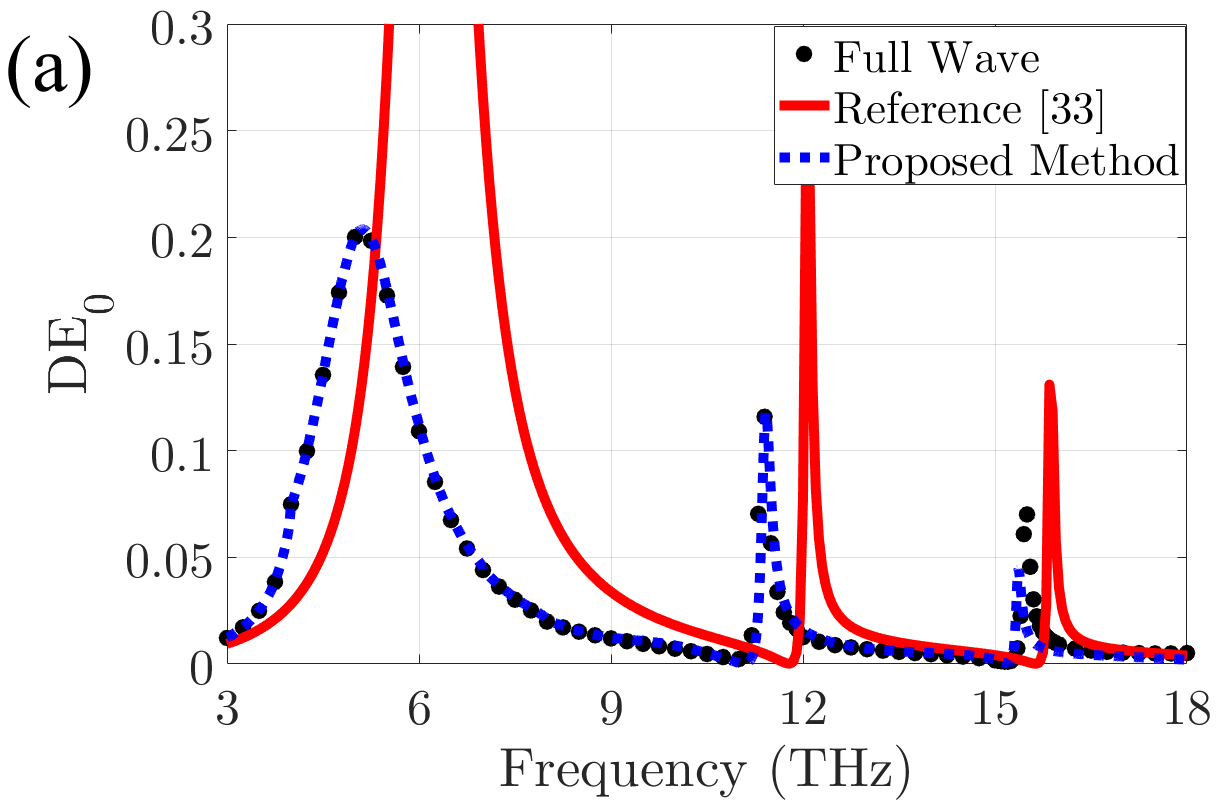}
\centering\includegraphics[width=\linewidth]{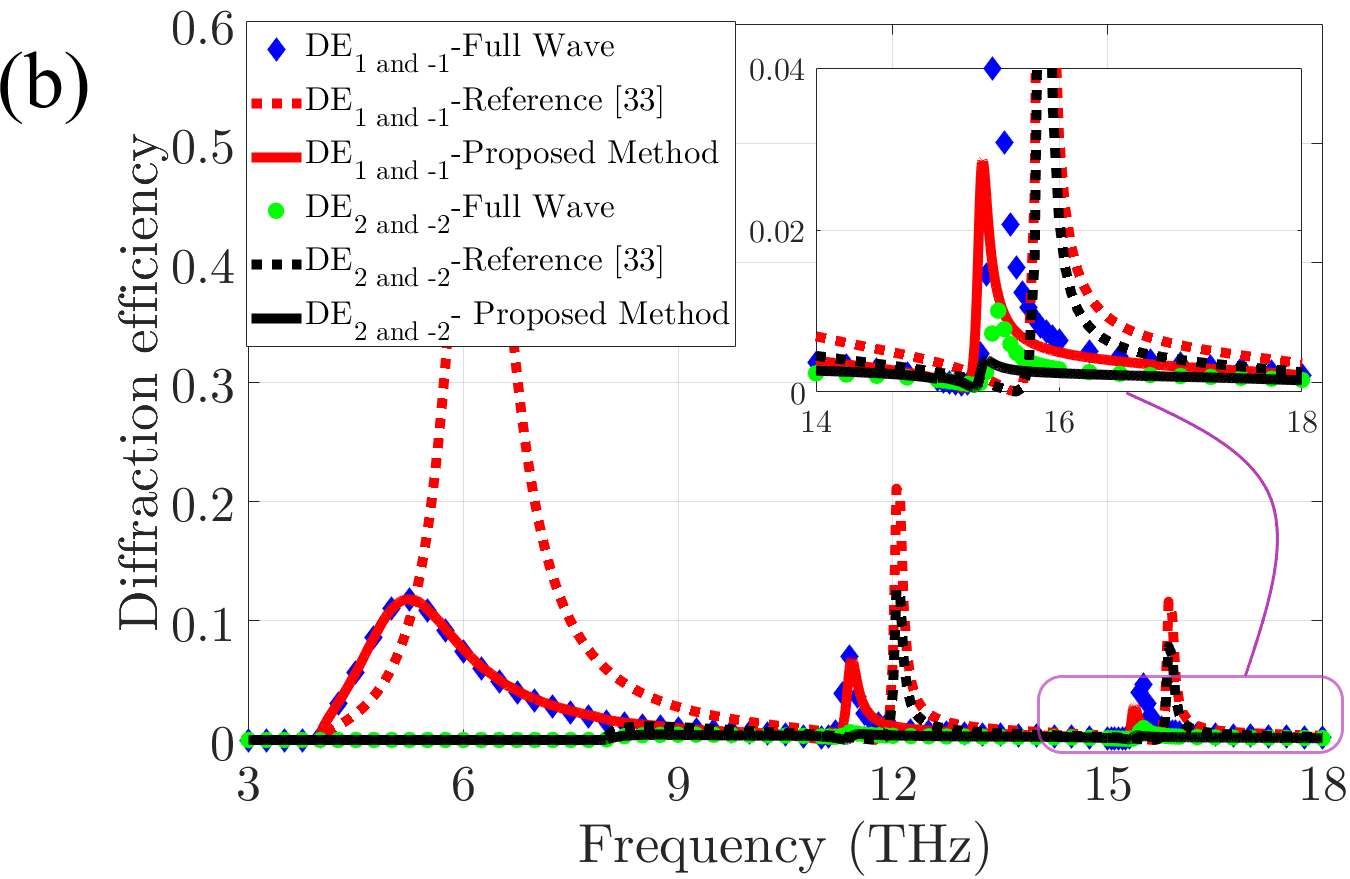}
\caption{The diffraction efficiency of (a) zeroth-order mode and (b) higher orders of  a periodic array of graphene ribbons with a width of  $w=15 \mu m$ and periodicity of $D=75 \mu m$. The graphene ribbons illuminated by a normal TM plane wave and the graphene parameters are assumed as $\tau=3 ps$ and $1.5 eV$.}
\label{ex1}
\end{figure}

\begin{figure*}
\centering\includegraphics[width=\linewidth]{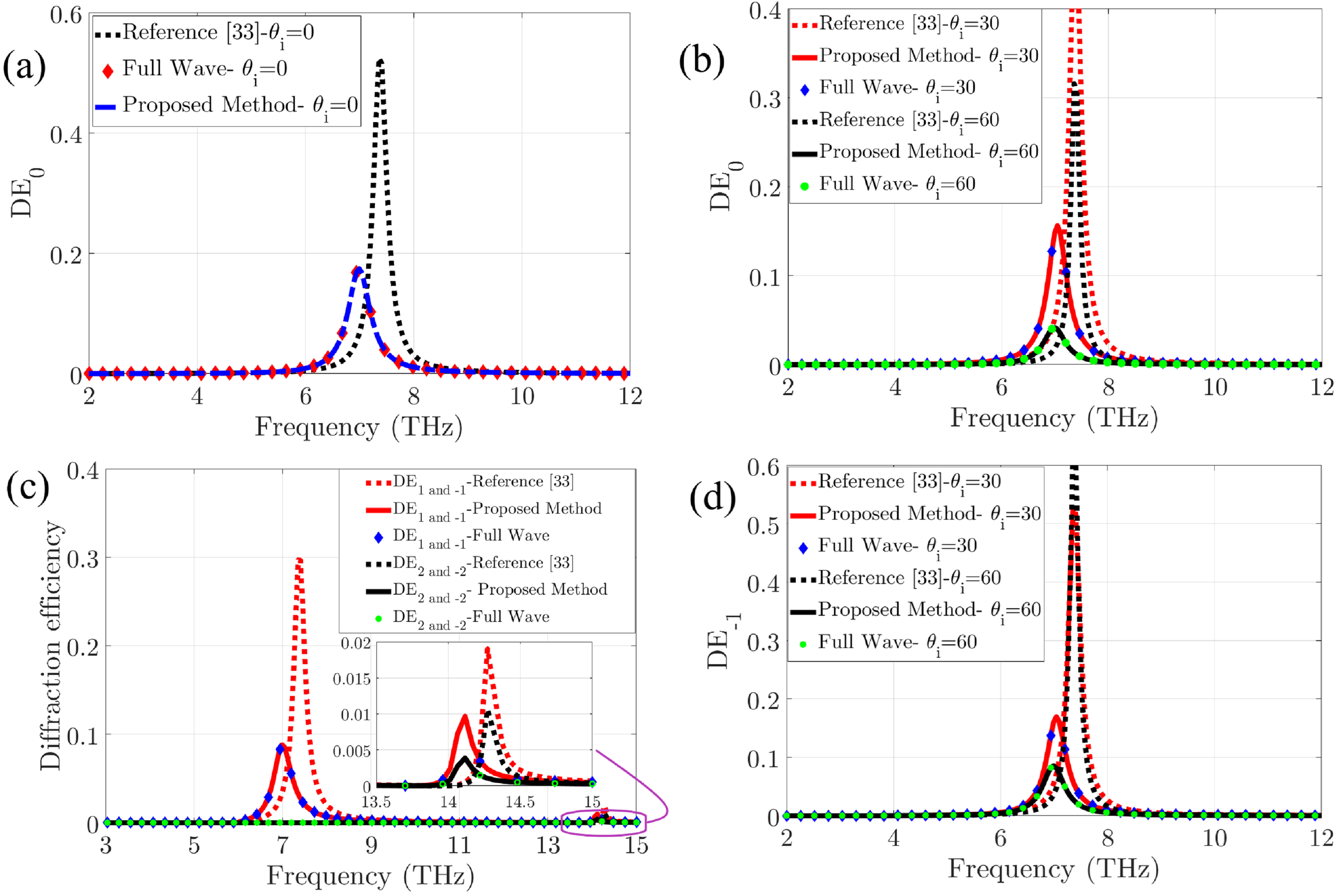}
\caption{Diffraction efficiency of a periodic array of graphene ribbons illuminated by an incident TM polarized wave. The graphene ribbons parameters are assumed as $D=50\mu m$, $w=5\mu m$, $E_F=0.7 eV$, and $\tau =2 ps$. (a) and (b) are the diffraction efficiencies of the zeroth-order mode for different incident angles. (c) The diffraction efficiencies of higher orders at normal incidence. (d)  $DE_{-1}$ for incident angles $\theta_i=30^\circ$ and $\theta_i=60^\circ$.}
\label{ex2}
\end{figure*}

In this section, some numerical examples are presented to verify the proposed analytical model and demonstrate its applicability. As the first example, we consider an array of graphene ribbons in free space.  Ribbon width and array constant are $w=15\mu m$ and $D=75\mu m$, respectively. The charge relaxation time and Fermi energy of graphene material are assumed to be $\tau = 3 ps$ and $E_F=1.5 eV$. The second example describes an array of graphene ribbons under TM oblique incidence. The parameters of the structure are $D=50\mu m$, $w=5\mu m$, $E_F=0.75 eV$, and $\tau =2 ps$. The diffraction efficiency of zeroth (specular) diffracted and higher diffracted orders are depicted in Fig.~\ref{ex1} and Fig.~\ref{ex2}. The results obtained are compared with the analytical method  in the subwavelength regime\cite{khavasi2014analytical},  and full wave simulations obtained using CST (Microwave Studio software). The latter is based on the finite integration technique (FIT) as is utilized as a reliable benchmark to verify the results. 

An excellent agreement is observed between the proposed model and the FIT results. Conversely, the subwavelength analytical method cannot predict the electromagnetic response of graphene ribbons. The methods introduced in \cite{khavasi2014analytical} uses a quasi-electrostatic approximation to derive the reflection coefficients of graphene ribbons. Hence, the operating wavelength must be larger than the periodicity of the structure (subwavelength limit). It is obvious from Fig.~\ref{ex1} and Fig.~\ref{ex2} that when the frequency increases, the accuracy of the previously introduced method decreases. By contrast, our method can analyze graphene ribbons in a wide range of frequencies.

\begin{figure}  [ht!]
\centering\includegraphics[width=\linewidth]{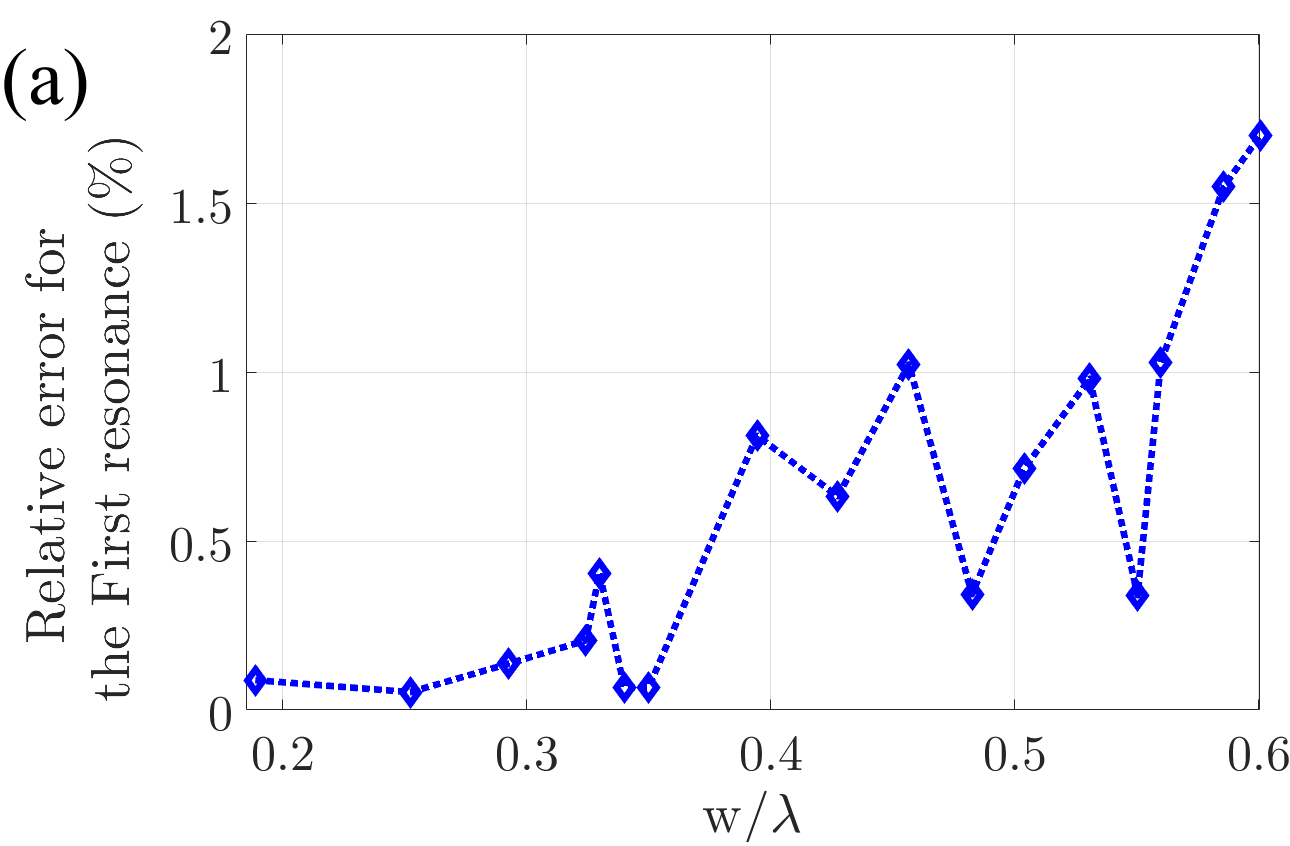}
\centering\includegraphics[width=\linewidth]{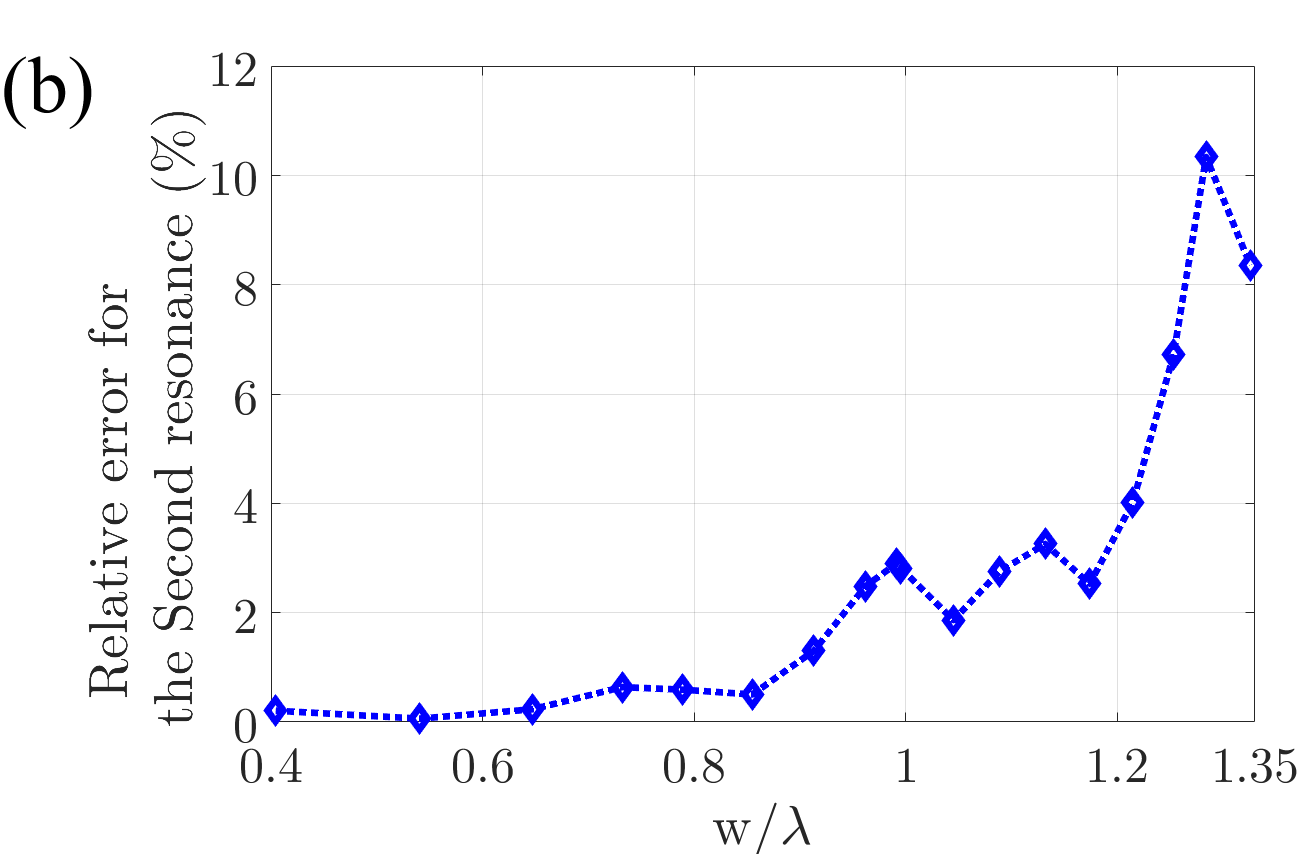}
\caption{The relative error for (a) the first and  (b) the second resonance frequency of array of graphene ribbons for different $w/\lambda$. Structure parameters are assumed as $D=100 \mu m$, $\theta_i=0$, $E_F=2 eV$, and $\tau =1.5 ps$.}
\label{Resonance_Freq}
\end{figure}

As for the limitations of the proposed method, we first note that while the array constant was arbitrary, the width of the ribbons was assumed to much smaller than the wavelength.   This allowed us to use the same eigenfunctions of a single, subwavelength ribbon under under quasi-static approximation. Therefore, we expect that when $w/\lambda$ increases, the proposed method becomes less accurate. To better clarify this point, the relative error for the frequency of the first and second resonances of the structure is plotted as function of $w/\lambda$ in Fig.~\ref{Resonance_Freq}. The parameters of graphene ribbons are $D=100\mu m$, $\theta_i=0$, $E_F=2 eV$, and $\tau =1.5 ps$. The graphene width is assumed to change between $5$ and $85 \mu m$. It can be seen from Fig.~\ref{Resonance_Freq}  that the relative error for the first resonance is relatively small: with practical values for graphene parameters, the first resonance frequency almost always occurs when $w/\lambda<1$, so that the proposed method can predict it accurately. As for the second resonance, the relative error for $w/\lambda=0.4$ is $0.2\%$, but rapidly increases with increasing width. For  $w/\lambda=1.3$ the error in predicting the second resonance frequency is $10\%$. This limitation does not pose a significant problem since graphene ribbons have usually been utilized near the first resonance \cite{khavasi2015design,abdollahramezani2015beam,rahmanzadeh2018adopting}.

\begin{figure}  [ht!]
\centering\includegraphics[width=\linewidth]{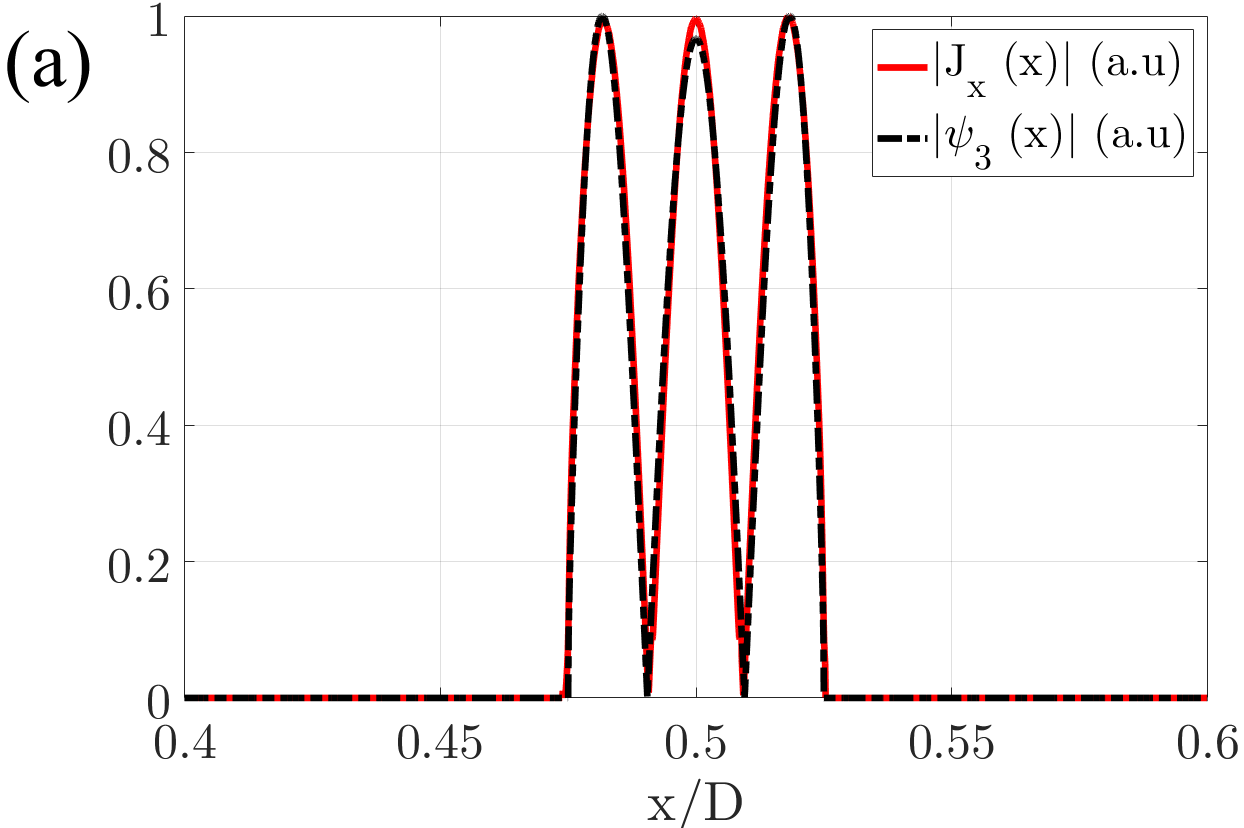}
\centering\includegraphics[width=\linewidth]{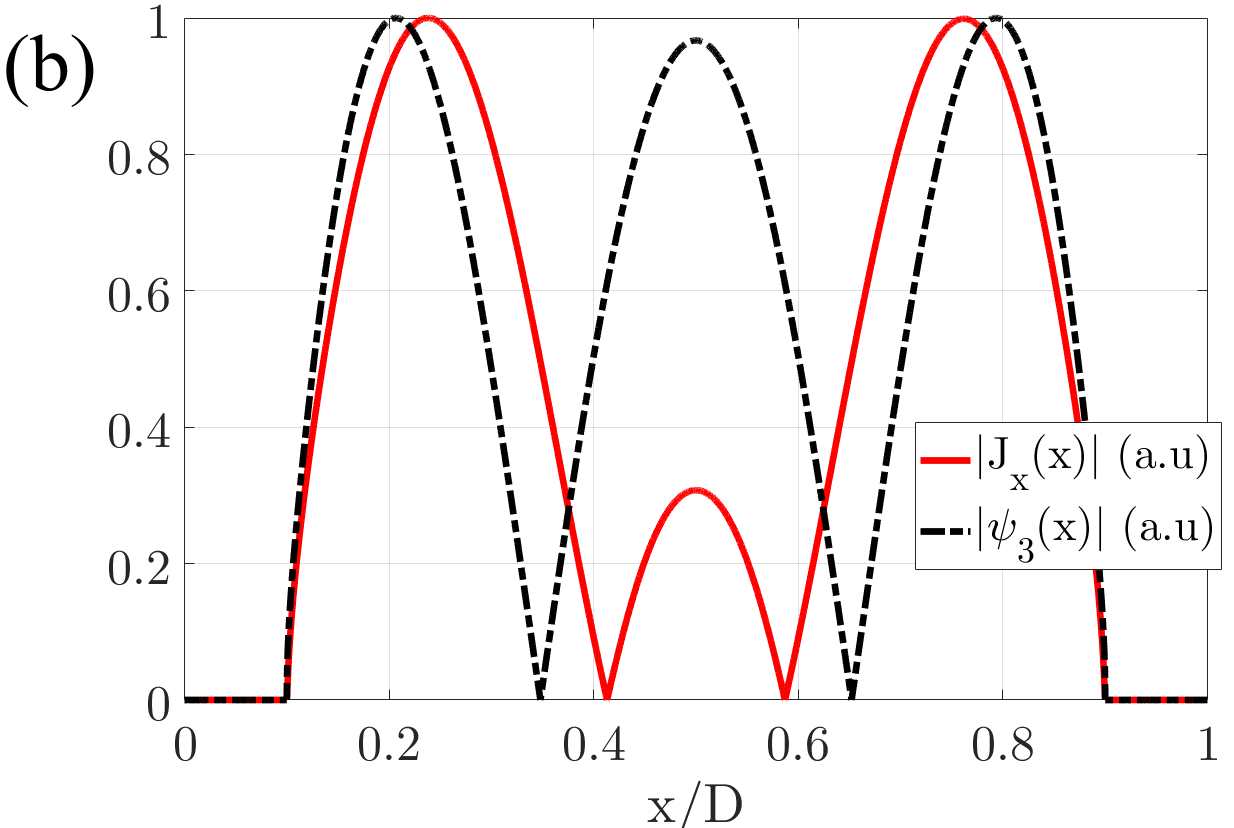}
\caption{The absolute value of current distribution on the graphene ribbons at the vicinity of the frequency of the second resonance and $|\psi_3(x)|$ for (a) $w/\lambda=0.4$ ($w=5 \mu m$) and (b) $w/\lambda=1.3$ ($w=80 \mu m$). The graphene ribbons parameters are assumed as $D=100 \mu m$, $\theta_i=0$, $E_F=2 eV$, and $\tau =1.5 ps$. Current distribution is obtained by CST Microwave Studio. }
\label{Surface_currents}
\end{figure}

The surface current distribution on a graphene ribbon is shown in Fig.~\ref{Surface_currents} at second resonance for $w/\lambda=0.4$ and $w/\lambda=1.3$. The results were   directly obtained from full-wave EM simulations. For comparison, the corresponding eigenfunction is also plotted on the same figure. It should be noted that, due to normal incidence, the even modes are absent and the second resonance corresponds to the eigenfunction $\psi_3(x)$. Good agreement is observed between results for $w/\lambda=0.4$, which implies that the eigenfunction of a single graphene ribbon well describes the profile of the induced surface current of graphene ribbons in an array. However, as can be seen in Fig.~\ref{Surface_currents}, when $w/\lambda$ is too large, these eigenfunctions cannot correctly describe the induced current.

\section{Conclusion}
 In this work, an analytical method was proposed to calculate the reflection coefficients of diffracted orders for an array of graphene ribbons. The method is based on integral equations governing the surface current density induced on graphene ribbons. Unlike previous studies, our method is valid even for array constants much larger than the wavelength of the incident wave. Closed-form and analytical expressions were presented for the diffraction efficiency of diffracted orders. The results obtained were in excellent agreement with those obtained from full-wave simulations. The model presented is fast, precise, and reliable and may be useful for devising various novel tunable devices in optics and photonics.

\bibliographystyle{IEEEtran}
\bibliography{main,Bibliography}

\end{document}